\documentstyle[12pt]{article}
\def\R1{\varepsilon_1}
\def\E8{\varepsilon_8}

\newcommand{\nn}{\nonumber}

\newcommand{\bea}{\begin{eqnarray}}
\newcommand{\eea}{\end{eqnarray}}
\newcommand{\be}{\begin{equation}}
\newcommand{\ee}{\end{equation}}
\newcommand{\bi}{\begin{itemize}}
\newcommand{\ei}{\end{itemize}}

\def\npb#1#2#3{    {\it Nucl. Phys. }{\bf B #1} (19#2) #3}

\def\plb#1#2#3{    {\it Phys. Lett. }{\bf B #1} (19#2) #3}
\def\prd#1#2#3{    {\it Phys. Rev. }{\bf D #1} (19#2) #3}

\def\prl#1#2#3{    {\it Phys. Rev. Lett. }{\bf #1} (19#2) #3}

\def\rmp#1#2#3{    {\it Rev. Mod. Phys. }{\bf #1} (19#2) #3}
\def\zpc#1#2#3{    {\it Zeit. f{\"u}r Physik }{\bf C #1} (19#2) #3}

\begin{document}

\begin{flushright}
        \small
        hep-ph/9906493\\
        June 1999
\end{flushright}

\begin{center}
\vspace{1cm}
{\large \bf The Contributions of Neutral Higgs Bosons to Charmless Nonleptonic
B Decays in MSSM}

\vspace{0.5cm}

\renewcommand{\thefootnote}{\fnsymbol{footnote}}
{ \bf Chao-Shang Huang{\footnote{\it csh@itp.ac.cn}} and Qi-Shu Yan{\footnote{\it yanqs@itp.ac.cn}}\\
Institute of Theoretical Physics, Academia Sinica.\\
P. O. Box 2735, Beijing 100080, P. R. China}

\vspace{.8cm}
{\large \bf abstract}
\end{center}
\vspace{0.5cm}

We investigate the contributions of neutral Higgs bosons to
nonleptonic transition $b \rightarrow q \bar{s} s, q=d,~s$ under the
supersymmetric context. Their effects to decay width and CP violation in
corresponding exclusive decays are explored. The anomalous dimension
matrices of the operators which have to be incorporated to include the
contributions of neutral Higgs bosons are given. We find that when
tan$\beta$ is large (say, 50) and neutral Higgs
bosons are not too heavy (say, 100 GeV), contributions of neutral Higgs penguin
can dominate electroweak penguin contributions, and for some processes, they
can greatly modify both decay width and CP asymmetry.

\newpage
\setcounter{page}{1}

The systematic study of the nonleptonic decays of B mesons
has been led long before the construction of B factories.
The CLEO report \cite{cleo} on the measurement of the charmless two-body
hadronic decays revived the theoretical interest in this subject. Among
the reported exclusive decay modes, $B \rightarrow K \eta'$ received
much more attention (for a review, see \cite{hych}), and according to \cite{ali,cheng},
four-quark-operator contribution is capable of explaining it under the theoretical framework of
SM. For the inclusive decay $B \rightarrow X_s \eta'$, it seems that new
mechanism or new physics should be introduced in order to explain the unexpected
large branching ratio \cite{incl}.

The effective Hamiltonian method is the fundamental tool to study the
charmless nonleptonic decays of B mesons. Starting from a full theory, after integrating
out the heavy freedom degree at a certain energy scale, the effective
Hamiltonian is obtained. For the standard model (SM), the heavy freedom
degrees
integrated out at $\mu_0=m_w$ are top quark, Z boson and W boson. For the SUSY models, in
order to be simple, the extra heavy freedom degrees integrated out
include all superpartners, though there are superpartners, such as the lightest
neutralino, which may be lighter than W boson, and there are superpartners,
such as the super-quarks of the first two generations, which may be much
heavier than
W boson. At this integrating out scale $\mu_0=m_w$, super-partners can
not only contribute to the effective Wilson coefficients, but
also introduce new
operators which will be shown below in the case of including
contributions of neutral Higgs bosons (NHB).

The works on the next-to-leading-order QCD-improved effective Hamiltonian
derived from SM were reviewed in \cite{efh}, where the problem on the scale
independence were studied. To present days, our knowledge of the dynamics of the
hadronic transitions are limited, and the approach employed in evaluating
hadronic matrix elements of operators is factorization.
There are two last papers which represent
the state of art of the systematic study of charmless hadronic two-body B decays.
Ref \cite{ali} concentrates on the
experimental test on the naive factorization ansatz, having classified
76 decay modes into five classes and analyzed the dependence of these decay
modes on the weak mixing matrix elements, form factors, decay constants,
QCD scale, and the effective number of color. Ref \cite{cheng} shows the
experimental data of $\rho^0 \pi^{\pm}$ and $\phi K^{\pm}$ cannot be
accommodated simultaneously by treating $N_c^{eff}(LL)=N_c^{eff}(LR)$ and,
by assuming
that $N_c^{eff}(LL)\simeq 2$ and $N_c^{eff}(LR) \sim 5$, finds the existed
experimental data can be explained well.

Nonleptonic decays have also been investigated in SUSY models
\cite{suhadr,suhadr1}.
Several papers  deal with the possible effects of SUSY to hadronic B
decays in the SUSY models with R parity violation \cite{rpvl}. The other
topic related with
the nonleptonic B decays of SUSY models is CP violation, there are a
vast amount of literatures \cite{subcp,sunl}.

It is well known that for charmless B hadronic decays with $\Delta B=1$
in which tree level contributions are suppressed by CKM  and color
symmetry, penguin contributions, among which gluonic ones play major
parts, always dominate.
In this letter we investigate the total contributions of penguins in
MSSM, including NHB penquins in addition to QCD penguins and electrweak
penguins. The motivation of this paper is based upon the fact found by
us in \cite{mtv},
where it is found that in SUSY models, when tan$\beta$ is large, NHB are
not heavy, and the mass splittings of stops and charginos are large,
contributions from exchanging NHB can greatly modified the branching ratio
and backward-forward asymmetry of $B \rightarrow X_s \l^{+} \l^{-},
l=\mu, \tau$.
It is a natural further step to explore their contributions to nonleptonic
B decays. Another motivation for our work is the striking large branching
ratio for the exclusive decay of $B\rightarrow K \eta'$ \cite{bepkex},
which has caused much theoretical efforts to assess the theoretical
prediction of SM and the validity of factorization approach \cite{bepkth,ali,cheng}.
Our results show that contributions of NHB can be larger than the sum of the
electroweak contributions from exchanging $\gamma$ and Z
bosons, and for some decay modes, they can modify both the decay widths
and CP asymmetries significantly.

We shall concentrate on the decay modes in which the quark level
process is $b\rightarrow q \bar{s} s, q=d,~ s$ in order to see the
contributions from NHB. The transition $b\rightarrow
q {\bar s} s$ is purely induced by FCNC penguin diagrams,
which can be divided into four classes, 1)those of exchanging gluon, 2) of exchanging photon, 3)
of exchanging Z, and 4) of exchanging Higgs neutral bosons.
Normally, contributions from the first class are expected to be large,
while those of the
second and the third are small because of $\alpha_{em}$/$\alpha_s$ and
$\alpha_2$/$\alpha_s$ respectively. In the SM, since suppressed by the ratio
of masses of quarks and masses of NHB, those of the fourth are also
negligible
, so only the contributions of the first are
considered. But many  papers \cite{ewp1} (for a review, see \cite{ewp2})
have shown that contributions from the second and the third classes
can play an important role in both the width decays and direct CP
violation in some decay modes,
and they can even play the major part in some processes, $B_s \rightarrow K^+ K^-$,
for instance.
Furthermore, in SUSY context, when tan$\beta$ is large and the masses of NHB
are not too heavy, contributions from exchanging neutral
Higgs bosons can be considerably large and therefore can greatly modify
theoretical predictions for some processes.
So in the letter, we consider the case of large tan$\beta$ and take into
account all contributions of the four classes.

The b transition $b\rightarrow q, q=d,~s;$ can be
described by the $\Delta B = 1$ effective Hamiltonian
\begin{eqnarray}
H_{\Delta B=1} = {4G_F\over \sqrt{2}}[\sum_{q=u,c} V_{q'b}V^*_{q'q}(c_1O_1 + c_2 O_2)
 - V_{tb}V^*_{tq}[\sum^{14}_{i=3} c_iO_i + \sum^{20}_{17} c_i O_i] +H.C.\;,
\end{eqnarray}
where $O_i$ are defined as
\begin{eqnarray}
O_1 = \bar q_\alpha \gamma_\mu L q'_\beta\bar
q'_\beta\gamma^\mu L b_\alpha\;&,&\;\;\;\;
O_2 = \bar q \gamma_\mu L q'\bar
q'\gamma^\mu L b\;,\nonumber\\
O_{3(5)} = \bar q \gamma_\mu L b \sum_{q'}
\bar q' \gamma_\mu L(R) q'\;&,&\;\;\;\;
O_{4(6)} = \bar q_\alpha \gamma_\mu L b_\beta \sum_{q'}
\bar q'_\beta \gamma_\mu L(R) q'_\alpha\;,\nonumber\\
O_{7(9)} ={3\over 2}\bar q \gamma_\mu L b \sum_{q'} e_{q'}\bar q'
\gamma^\mu R(L)q'\;&,&\;\;
O_{8(10)} = {3\over 2}\bar q_\alpha \gamma_\mu L b_\beta \sum_{q'}
e_{q'}\bar q'_\beta \gamma_\mu R(L) q'_\alpha\;,\nonumber\\
O_{11(13)} = \bar q R b \sum_{q'} \bar q' L(R) q'\;&,&\;\;\;\;
O_{12(14)} =\bar q_\alpha R b_\beta \sum_{q'} \bar q'_\beta L(R) q'_\alpha\;,\nn\\
O_{17} =\bar q \sigma^{\mu \nu} R b \sum_{q'} \bar q' \sigma_{\mu
\nu} R q'\;&,&\;\;\;\;
O_{18} =\bar q_\alpha \sigma^{\mu \nu} R b_\beta \sum_{q'} \bar q'_\beta \sigma_{\mu \nu}
R q'_\alpha\;,\nn\\
O_{19} = \frac{g_s}{8 \pi^2} m_b \bar q \sigma^{\mu \nu} \lambda^a
R b G^a_{\mu \nu} \;&.&\;\;\;\; O_{20} = \frac{e}{8 \pi^2} m_b
\bar q \sigma^{\mu \nu} R b F_{\mu \nu} \;
\end{eqnarray}

The first two operators are induced by tree diagrams,
$O_{i=3,\cdots,6}$ by gluonic penguins, $O_{i=7,\cdots,10}$ by
$\gamma$ and Z penguins, $O_{i=11,\cdots,14}$ and $O_{i=17,18}$ by
neutral Higgs penguins, and $O_{19,20}$ by chromomagnetic and
magnetic penguins\footnote{In order to make the minimal revisions
we use the labels of operators as those in the original version of
the paper. Comparing with those in refs.~\cite{ch,chw,hz,hk},
$O_{17},O_{18},O_{19},O_{20}$ correspond to
$O_{15},O_{16},O_{8g},O_{7\gamma}$ respectively.}.

The boundary conditions of the first ten Wilson coefficients and
those of $O_{19,20}$ for the solutions of the renormalization
group equations are given in \cite{efh}, here we provide those of
$O_{11,\cdots,14,17,18}$ in the large $\tan\beta$ case.
\begin{eqnarray}
C_{O_{11}}(m_w)&=&\delta_{sq'} \frac{\alpha_2}{16 \pi} \sqrt 2 tan^3\beta m_b m_s (\frac{\cos^2\alpha}{m^2_{H^0}}+\frac{\sin^2\alpha}{m^2_{h^0}}+\frac{1}{m^2_{A^0}})
\left \{\sum_{i=1}^{2}\frac{m_{\chi_i}}{m_w} \left
[ \right.\right. \nn\\
& & U_{i2} V_{i1} f(x_{\chi_i w},x_{{\tilde q}w})
-\sum_{k=1}^{2} U_{i2} T_{k1} \left (V_{i,1} T^*_{k1}\right .\nn\\
&&\left. \left. \left. -\frac{m_t}{\sqrt 2 \sin\beta} V_{i2} T^*_{k2} \right) f(x_{\chi_i w},x_{{\tilde t_k} w})\right] \right \}. \nn\\
C_{O_{12}}(m_w)&=&0;\nn\\
C_{O_{13}}(m_w)&=&\delta_{sq'} \frac{\alpha_2}{16 \pi} \sqrt 2 tan^3\beta m_b m_s (\frac{\cos^2\alpha}{m^2_{H^0}}+\frac{\sin^2\alpha}{m^2_{h^0}}-\frac{1}{m^2_{A^0}})
\left \{\sum_{i=1}^{2}\frac{m_{\chi_i}}{m_w} \left
[\right. \right.\nn\\
& &U_{i2} V_{i1} f(x_{\chi_i w},x_{{\tilde q}w})
-\sum_{k=1}^{2} U_{i2} T_{k1} \left (V_{i,1} T^*_{k1}\right. \nn\\
&&\left. \left. \left. -\frac{m_t}{\sqrt 2 \sin\beta} V_{i2} T^*_{k2}\right ) f(x_{\chi_i w},x_{{\tilde t_k} w})\right ] \right \}. \nn\\
C_{O_{i}}(m_w)&=&0, ~i=14,17,18,
\end{eqnarray}
where
\begin{eqnarray}
f(x,y)=\frac{-1}{x-y}(x \log x-y \log y),\;\;
x_{ij}=\frac{m_i^2}{m_j^2}\;\;
\end{eqnarray}

The conventions can be found in \cite{mtv}. To reach these, we
have neglected some terms contributing minor. It is interesting to
mention that the terms proportional to tan$^3\beta$ only come from
the one Feynman diagram where firstly virtual NHB are emitted from
b-quark line and decay into strange quark pair, then followed with
flavor change caused by loops exchanging charginos and stops.
While for virtual NHB decaying to down quark pair, because of the
smallness of $m_d$/$m_s$ the contributions are negligibly small.
To up-type quark pair, SUSY contributions at most be  proportional
to tan$^2\beta$ and consequently also not important compared to
those for decaying to strange quark pair. This is the reason why
we only pay attention to the transition $b\rightarrow q \bar{s}
s$.

$C_{O_{11,13}}$ are proportional to $x_{\chi_i w}\log x_{\chi_i
w}$. Therefore, contributions of SUSY depend on not only masses
difference between stops, but also those between charginos. In
order to escape the GIM cancellation and to increase the
contributions of SUSY, both large mass splittings of stops and
charginos and large chargino masses are needed. In contrast with
the process $b\rightarrow s \gamma$ and the mixing of $B^0$
system, the smaller the mass of the lighter charginos is, the
larger the contributions of SUSY.

For the first ten Wilson coefficients, we use next-leading-order
QCD corrected renormalization group equations
\cite{efh}\footnote{We do not include the mixing of $O_{11,12}$
onto $O_{3,\cdots,10}$ which has been given in ref.~\cite{hk}.We
also do not include the mixing of $O_{13,14,17,18}$ onto
$O_{19,20}$ which is given first in ref.~\cite{bghw} and verified
in ref.~\cite{hk}.}, while for the last six ones (i.e.,
$O_{11,\cdots,14,17,18}$), we use leading-order ones since we
carried out calculations of the relevant anomalous dimension
matrices only at one-loop\footnote{We would like to thank Dr. C.D.
L\"u for his collaboration on the calculations. A complete
analysis of mixing of the operators will be published else
where.}. The anomalous dimension matrices of the last six
operators (i.e., $O_{11,\cdots,14,17,18}$) can be divided into two
distangled groups
\begin{eqnarray}
\gamma^{(L)}=\begin{tabular}{c|cccc} &$O_{11}$&$O_{12}$\\\hline
$O_{11}$&$-16$&0\\ 
$O_{12}$&-6&$2$ 
\end{tabular}
\end{eqnarray}
and
\begin{eqnarray}
\gamma^{(R)}=\begin{tabular}{c|cccc}
&$O_{13}$&$O_{14}$&$O_{17}$&$O_{18}$\\\hline
$O_{13}$&$-16$&0&1/3&-1\\
$O_{14}$&-6&$2$&-1/2&-7/6\\
$O_{17}$&16&-48&$16/3$&0\\
$O_{18}$&-24&-56&6&$-38/3$
\end{tabular}
\end{eqnarray}

Our calculation shows that $C_{O_{12,\cdot,14,17,18}}(m_b)$ are
much smaller than $C_{O_{11}}(m_b)$, so it is appropriate to
neglect them in our numerical analysis.

As given in \cite{efh}, the effective coefficients $a_i$ are dependent of $N_c^{eff}$.
In this note, we do not examine the dependence and limit ourself to a typical value of
$N_c^{eff}$, 3, in order to see the contributions of NHBs. For $N_c^{eff}=3$ and
transition $b\rightarrow s$, the first ten effective coefficients given from
Wilson coefficients in SM are \cite{ali}
\begin{equation}
\begin{array}{ll}
a_1^{eff} =1.05 (1.05),&a_2^{eff} =0.053 (0.053),\\
a_3^{eff} =48 (48),&a_4^{eff} = -439-77 i (-431-77 i),\\
a_5^{eff} =-45 (-45),&a_6^{eff} = -575-77 i (-568-77 i),\\
a_7^{eff} =0.5-1.3 i (0.5-1.3 i),&a_8^{eff} = 4.6-0.4 i (4.6-0.4 i),\\
a_9^{eff} = -94-1.3 i (-94-1.3 i),& a_{10}^{eff} = -14-0.4 i(-14-0.4 i)
\end{array}\\
\label{eq:input1}
\end{equation}
while for transition $b\rightarrow d$, they are
\begin{equation}
\begin{array}{lll}
a_1^{eff} = 1.05 (1.05),&a_2^{eff} =0.053 (0.053),\\
a_3^{eff} = 48 (48),&a_4^{eff} = -412- 36 i (-461-124 i),\\
a_5^{eff} =- 45 (-45),&a_6^{eff} = -548-36 i (-597-124 i),\\
a_7^{eff} = 0.7-1 i (0.3-1.8 i),&a_8^{eff} = 4.7-0.3 i (4.5-0.6 i),\\
a_9^{eff} = -94-1 i (-95-1.8 i),& a_{10}^{eff} = -14-0.3 i (-14-0.6 i)
\end{array}\\
\label{eq:input2}
\end{equation}

These complex figures incorporate contributions of both the strong phases and weak phases.
For the last 8 coefficients, $10^{-4}$ should be multiplied, and the numbers in the brackets
refer to the conjugate processes $\bar b \rightarrow
\bar s s \bar s$ and $\bar b \rightarrow \bar d s \bar s$.

The input parameters used in our numerical calculations are as follows.

Wolfenstein parameters \cite{wolf} for the CKM matrix are given as $A=0.81$,
$\lambda=0.2205\pm0.0018$ \cite{fit1}, $\rho=0.12$ and $\eta=0.34$ \cite{fit2}.
The scale for the running masses of quarks is set to 2.5 Gev, and the corresponding
masses are given in Table I in \cite{ali}. We use the BSW approach to
evaluate the form factors \cite{bsw}, and the form factors at zero momentum
can be found in the Table II in~\cite{ali}. For the reason that the form factors
are not sensitive to the pole masses since only small extrapolations from $Q^2$ = 0 are involved
in the decays $B\rightarrow h_1 h_2$, we set the pole mass to 5.4, which is given in Table III in~\cite{ali},
in calculating the form factors . Decay constants are given in the Table V in~\cite{ali}.
The mixing angles of the flavor SU(3)-octet and singlet components are set
to $\theta_8=-21.2^\circ$ and $\theta_0=-9.2^\circ$ \cite{angle}.
For more detailed information
on conventions, please refer \cite{ali}.

We calculate $C_{O_{11}}$ and find that it can reach $0.05$. So
neutral Higgs bosons can dominate the contributions of the
electro-weak penguins for some processes. In calculations the
relevant ten supersymmetric parameters are taken as
\begin{equation}
\begin{array}{llll}
M_2 =300~GeV,&\mu = 300~GeV,&m_{A^0} = 80~GeV,&m_{\tilde u}=500~GeV,\\
m_{R} = 300~GeV,&m_{L} = 300~GeV,&A_t = 300 GeV,& tan\beta=50\\
&\psi_{\mu}=\pi& \psi_{A_t}= 0.3&,
\end{array}
\end{equation}
which satisfy the constraint
on the SUSY phases arising from EDMs through the Barr-Zee
mechanism~\cite{bz,ckp} as well as the constraint from $b\rightarrow s \gamma$~\cite{mtv}.

we get $C_{O_{11}}(m_b)=-0.051 - 0.027 i$, and consequently $a_{11}^{eff}= -0.051-0.027 i$ and
$a_{12}^{eff}=-0.017-0.009 i$. We note that the parameters (9) satisfy the
constraint from $b\rightarrow s \gamma$.
For the first ten Wilson coefficients, we find that the contributions of SUSY
with the above parameters can be safely neglected, because they only
modify the Wilson coefificients in few percent in most of part of the
parameter
space. So we shall still use eqs. (\ref{eq:input1}) and (\ref{eq:input2})
for the sake of simplicity.
The complex phase of $C_{O_{11}}$ is originate from the SUSY CP phases,
if $\psi_{A_t}= 0$, the corresponding $a_{11}^{eff}= -0.055$,
no imaginary part appears.

To evaluate the relevant transition matrix elements, we will use the
naive factorization assumption.
In the spectator model, several formula on decay amplitudes
are listed below
\begin{eqnarray}
M( \bar B^0 \to \bar K^0 \eta ^{(\prime)} )  &=&-\frac{G_F}{\sqrt{2}} M^{B\rightarrow \eta^{(\prime)}}_{sss}(K) V_{tb}V_{ts}^*\left [
a_4-\frac{1}{2} a_{10} + \right. \nn\\
&&\left . (a_6-\frac{1}{2} a_8) R_5\right]
-\frac{G_F}{\sqrt{2}} M^{B \rightarrow K}_{sss}
\left \{ V_{ub}V_{us}^*  a_1 +\right. \nn\\
&&V_{cb}V_{cs}^*  a_1 \frac{ f_{\eta ^{(\prime)}}^c} {f_{\eta^{(\prime)} }^u}-
V_{tb}V_{ts}^* \left
[ 2a_3 -2a_5 +\frac{1}{2}(a_{9}-a_7)
+(a_3-a_5 \right. \nn\\
&&+a_9-a_7)\frac{f_{\eta^{(\prime)}}^c}{f_{\eta^{(\prime)}}^u}+
\left( a_3+ a_4-a_5+
\frac{1}{2}(a_7-a_9-a_{10}-a_{12})+\right .\nn\\
&&\left.\left.\left.(a_6-\frac{1}{2}(a_8-a_{11})) R^{(\prime)}_6 \right )
(\frac{ f_{\eta ^{(\prime)} }^s} {f_{\eta ^{(\prime)}
 }^u}-1)\right] \right \} . \nn\\
M( \bar B^0 \to \bar K^0 \phi )  &=&-\sqrt{2} G_F M^{B \rightarrow
K}_{sss}(\phi) V_{tb}V_{ts}^*\left [
a_3+a_4+a_5 \right. \nn\\
&&\left. - \frac{1}{2}(a_7+a_9+a_{10} +a_{12}) \right]
\end{eqnarray}

Where
\begin{eqnarray}
M^{B \rightarrow P}_{q_1 q_2 q_3}(P'(V))&=&\langle P'(V) \vert {(\bar{q_1} q_2)}_{V-A} \vert 0
\rangle \langle P \vert {(\bar{q_3} b)}_{V-A} \vert \bar B^0 \rangle,\nn \\
R_5&=&\frac{2 m_K^2}{(m_b-m_u)(m_u+m_s)},\nn \\
R^{(\prime)}_6&=&\frac{2 m_{\eta^{(\prime)}}}{(m_b-m_s)(m_s+m_s)}.
\end{eqnarray}
and the superscript "eff" of $a_i^{eff}$ has been omitted for the sake of
simplicity.

Compared with formula given in \cite{ali}, we have taken into account
the contributions of NHB's by adding $a_{11}$, $a_{12}$ and $a_{13}$ in
these amplitudes of decays.
We have not included $W$-exchange, $W$-annihilation and spacelike penguin
contributions because the W-exchange and W-annihilation contributions are
negligible due to helicity suppression (as well as color suppression for
W-exchange) and we do not have a reliable method to estimate the
spacelike penguins\cite{cheng}. We have used the equations of motion
and Fierz rearrangement to calculate the contributions of scalar operators.
For other formula, it is easy to derive when compared with the tables given
in the appendices in \cite{ali,cheng}.
Our numerical analysis about the effects of NHB are based on these formula.

For charged $B^{\pm}$, $A_{CP}$ is defined as
\begin{eqnarray}
A_{CP}&=&\frac{\Gamma(B^- \rightarrow f^-)-\Gamma(B^+ \rightarrow f^+)}{\Gamma(B^- \rightarrow f^-)+\Gamma(B^+ \rightarrow f^+)}
\end{eqnarray}
and it reflects the magnitude of the direct CP violation.
For the cases
in which both $B^0$ and $\bar B^0$ decay to the same final states, $A_{CP}$
is defined as
\begin{eqnarray}
A_{CP}&=&\frac{1-|\epsilon|^2-2 Im (\epsilon x_B)}{(1+|\epsilon|^2) (1+x_B^2)}
\end{eqnarray}
where $\epsilon=\frac{q}{p} \frac{{\bar A}}{A}$, a quantity reflecting effects
from both the mixing and the direct CP violations. We
omit the analysis on the CP violation for decays, $B^0 \rightarrow V V'$, and
$\bar B^0 \rightarrow \bar K^{*0} K^0, K^{*0} \bar K^0$, in order to
simplify
the calculations.
$x_B$ is set to 0.723 as given in \cite{pdg}. We do not introduce new CP origin
in the mixing of the neutral B meson.

Our result is shown in Table 1. A number of observations are in order.

1. Compared to the SM, the decay widths  for most modes corresponding to the transition
$b\rightarrow q \bar s s$ increase. For example, for $B^- \rightarrow K^- \eta'$, contributions
of NHB increase the branching ratio up to 25$\%$ and for $B^- \rightarrow K^-
\eta$, they increase up to 80$\%$. For $B\rightarrow K(K^*) \phi$ ($\bar B^0 \rightarrow \pi^0 \eta'$)
because of the cancellations between $a_{12} (a_{11})$ and $a_4 (a_6)$, the decay widths decrease.
We see that the contributions of NHB do modify the branching ratios for all
charmless decay modes whose amplitudes are governed by penguins for
transition $b\rightarrow q \bar{s} s$, though for most nonleptonic decays
gluon penguins dominate.
Compared with the semileptonic decays $b\rightarrow s l^+ l^-$, where
branching ratios and forward-backward asymmetry can be greatly
modified \cite{mtv} by the effects of NHB, for most charmless nonleptonic
decay modes their effects are soften by the large values of the
coefficients $a_1, a_4, a_6$.

2. For those decay modes where electroweak penguins play an important role, such as
$B\rightarrow K^{*}\eta$, $\bar B^0 \rightarrow \bar{K}^{*0} (K^0) \eta$,
$\bar B^0 \rightarrow \pi^0 \eta$,$B^0\rightarrow \eta (\eta') \eta (\eta')$, and $\bar B^0 \rightarrow \omega^0
\eta'$, the contributions of NHB can enhance the branching ratios by several
handreds percent.

3. For the modes where there are subtle cancellations between contributions in SM,
such as $B \rightarrow K^{*0} K$, $\bar B^0 \rightarrow K^{*0} \eta'$,
and $\bar B^0 \rightarrow \rho^0 \eta'$,
the contributions of NHB can enhance the branching ratios by a factor of a few tens
 and greatly modify the corresponding CP asymmetry.

 In the effective Hamiltonian method
to study hadronic decays the largest theoretical uncertainty arises from the calculations of  hadronic
matrix elements of local operators. For energetic two-body decays of B mesons  factorization should be a good
approximation because of color transparency  ~\cite{bj} and non-factorization effects may be parameterized
by $N_c^{eff}$. In the factorization framework, in addition to the uncertainty due to non-factorization effects,
  the other uncertainties come from the input parameters, such as form factors,
decay constants, quark masses, CKM matrix elements, the scale at which Wilcon coefficients is calculated, etc.
The different choices of input parameters can lead to an uncertainty of from a few percent to several handreds
percent (dependent on modes), as can be seen by comparing the results for the
 same $N_c^{eff}$ between ref.\cite{ali} and ref.\cite{cheng}.
The uncertainty due to different $N_c^{eff}$ can also reach several handreds percent. Therefore, it is difficult
to disentagle the non-standard physics for most of modes in Table 1. However, by comparing our result and the results
given in ~\cite{ali} and \cite{cheng}, it can be drawn that \\
A. compared to SM, a significantly large enhancement of the branching retio for $B\rightarrow K^* \eta' $ in MSSM is definitely
existed; and\\
B. the branching ratio for $B\rightarrow \rho^0 \eta' $ ($\bar B^0\rightarrow K^{*0} \bar K^0$) is enhanced by a factor of 40 (25) compared to SM
for the same $N_c^{eff}$ and   larger than all values corresponding to different $N_c^{eff}$ in SM given in \cite{ali,cheng}. Thus,  it could
be possible to disentagle MSSM from SM by measuring the branching ratio if tan$\beta$ is large and NHBs are not heavy.

In summary, we have analyzed the contributions of NHB to charmless
nonleptonic two-body B decays. Our analysis shows that at some regions of
the parameter space in
MSSM, NHBs can dominate the contributions of electroweak
penguins and significantly modify both decay widths and CP asymmetry of
the decay modes which are due to the transition $b\rightarrow q \bar{s}
s$.
Our analysis has not include CP phase of SUSY origin in mixing.
After taking into the complex contributions of SUSY to the mixing, CP
asymmetry of most of neutral B meson decay modes could be modified.

>From the above analysis, we know that  the effects of NHB, for most charmless nonleptonic
decay modes,  are soften by the large values of the
coefficients $a_1, a_4, a_6$. But their effects can show up when there are
subtle cancellations. Therefore, although there are huge theoretical uncertainties, the enhancements
for some modes, such as $B\rightarrow \rho^0 \eta' $ and $\bar B^0\rightarrow K^{*0} \bar K^0$, exceed
the theoretical uncertainties significantly so that it would be possible
to search new physics effects by measuring these decay modes.

\section*{Acknowledgment}
C.S. Huang would like to acknowledge KIAS where part of the letter was
written for the warm hospitality. Q.S. Yan  would like to thank the
helpful discussions with Dr. O. Vives, Dr. C.D. L\"u and Dr. K.C. Yang.
This work was supported in part by the National Natural Science Foundation
of China and partly supported by CCAST.

\begin{table}
\vspace*{1.5cm}
\caption{Branching ratios in unit of $10^{-6}$,and $CP$
asymmetry in unit of $\%$.}
\vspace*{0.5cm}
\begin{center}
\begin{tabular}{lcccccc}
$N_c=3$ & \multicolumn{2}{c}{Branching~Ratio} & \multicolumn{2}{c}{CP~Asymmetry} \\
Decay Modes&SM&SUSY&SM&SUSY\\\hline
 $B^-\rightarrow K^-\eta$&1.8&3.27&-6.1&-0.16\\
 $B^-\rightarrow K^-\eta^{'}$&25.2&31.9&4.5&5.58\\
 $B^- \rightarrow K^- \phi$&13.8&8.11&2.0&-4.8\\
 $B^-\rightarrow K^{*-} \eta$&1.50&6.64&5.4&3.3\\
 $B^-\rightarrow K^{*-} \eta^{\prime}$&0.89&6.71&23.7&-5.4\\
 $B^- \rightarrow K^{*-} \phi$&6.95&4.43&2.0&-3.7\\\hline
 $B^-\rightarrow K^- K^0$&0.67&0.82&-12.3&-9.8\\
 $B^- \rightarrow \pi^- \eta$&2.10&2.14&-14.0&-13.3\\
 $B^- \rightarrow \pi^- \eta'$&2.16&2.56&-13.0&-11.7\\
 $B^- \rightarrow K^0 K^{*-}$&0.46&0.69&-14.5&-9.2\\
 $B^- \rightarrow K^{*0} K^-$&0.0004&0.01&-46.7&6.2\\
 $B^- \rightarrow \rho^- \eta$&6.16&6.89&-3.4&-3.1\\
 $B^- \rightarrow \rho^- \eta'$&7.41&6.99&-11.0&-11.3\\
 $B^- \rightarrow K^{*-} K^{*0}$&5.73&3.77&-14.5&-22.1\\\hline
 $\bar{B}^0\rightarrow \bar{K}^{0}\eta$ &1.6&3.1&27.8&37.5\\
 $\bar{B}^0\rightarrow \bar{K}^{0}\eta^{'}$&24.1&30.5&29.4&33.5\\
 $\bar B^0\rightarrow \phi \bar K^0$&13.3&7.8&29.3&15.1\\
 $\bar B^0\rightarrow \bar K^{*0} \eta$&1.43&6.14&30.0&41.4\\
 $\bar B^0\rightarrow \bar K^{*0} \eta^{\prime}$&0.43&6.4&24.6&37.2\\
 $\bar B^0\rightarrow \bar K^{*0} \phi$&6.69&4.27&---&---\\\hline
 $\bar B^0\rightarrow \bar K^0 K^0$&0.65&0.79&33.2&35.6\\
 $\bar B^0 \rightarrow \pi^0 \eta$&0.32&0.71&31.3&38.8\\
 $\bar B^0 \rightarrow \pi^0 \eta'$&0.12&0.019&26.8&-70.3\\
 $\bar B^0 \rightarrow \eta \eta'$&0.039&0.11&24.0&39.0\\
 $\bar B^0 \rightarrow \eta \eta$&0.13&0.35&24.0&39.0\\
 $\bar B^0 \rightarrow \eta' \eta'$&0.024&0.08&-4.3&-27.2\\
 $\bar B^0\rightarrow K^0 \bar K^{*0}$&0.44&0.66&---&---\\
 $\bar B^0\rightarrow K^{*0} \bar K^0$&0.0004&0.01&---&---\\
 $\bar B^0 \rightarrow \rho \eta$&0.015&0.0076&34.2&-34.7\\
 $\bar B^0 \rightarrow \rho \eta'$&0.0028&0.11&63.1&27.3\\
 $\bar B^0 \rightarrow \omega \eta$&0.046&0.048&-10.5&-56.3\\
 $\bar B^0 \rightarrow \omega \eta'$&0.019&0.11&-57.9&36.6\\
 $\bar B^0\rightarrow \bar K^{*0} K^{*0}$&5.52&3.63&---&---\\
 \end{tabular}
 \end{center}
\end{table}

\end {document}